\documentstyle[pre,aps,twocolumn]{revtex}
\begin{document}
\draft
\tighten

\title{Time evolution of an anharmonic oscillator interacting
       with a squeezed bath}

\author{G. M. D'Ariano\cite{dar}}
\address{Dipartimento di Fisica ``Alessandro Volta'', Universit\`a di Pavia,
         Via A. Bassi 6, I--27100 Pavia, Italy}
\author{M. Fortunato\cite{for}}
\address{%
Dipartimento di Fisica, Universit\`a di Roma ``La Sapienza'',
         P.le A. Moro 2, I--00185 Roma, Italy \\ {\rm \ } \\ {\rm and} }
\vskip -.3cm
\author{P. Tombesi\cite{tom}}
\address{ Dipartimento di Matematica e Fisica, Universit\`a di Camerino,
          Via Madonna delle Carceri, I--62032 Camerino, Italy}
\date{Received \today}
\maketitle
\widetext
\begin{abstract}
The evolution of a single mode of the electromagnetic field interacting
with a squeezed bath in a Kerr medium is considered. The solution
of the corresponding master equation is given numerically. It is argued that
the creation of a superposition state (Schr\"odinger's cat) is better achieved
in presence of a squeezed reservoir than of a thermal one.
\end{abstract}
\pacs{PACS number: 42.50.Dv}
\narrowtext
\section{Introduction}
\label{intro}
The possibility of detecting quantum interference effects between
macroscopically distinguishable states---eigenstates of a
variable corresponding to macroscopically different eigenvalues---has
received much attention in the last few years
\cite{kn:yusto,kn:milb,kn:mto,kn:tome,kn:meto,kn:kenw,%
kn:mimto,kn:mil,kn:brisu,kn:buz}. Such interest is motivated by
one of the fundamental problems in quantum mechanics: to determine
whether quantum-mechanical features can be observed in macroscopic objects
\cite{kn:legg}. Unfortunately, the detection of a macroscopic superposition is
very difficult, due to the unavoidable coupling with the environment and by
the consequent dissipation \cite{kn:zur}. The severity with which dissipation
destroys the quantum coherence at macroscopic level has been
discussed by several authors \cite{kn:miw,kn:caleg,kn:wamil,%
kn:sawa,kn:miho,kn:miwa,kn:dami}.

Since the original proposal by Yurke and Stoler \cite{kn:yusto},
who have shown that a coherent state propagating through a Kerr
medium evolves, under suitable conditions, into a quantum
superposition of two coherent states which are 180$^\circ$
out of phase with each other, many attempts have been made
in order to preserve this superposition with detectable effects as, for
example, the interference fringes at the output of a homodyne detector.
Mecozzi and Tombesi \cite{kn:mto,kn:tome,kn:meto} have considered a
dissipation model corresponding to a beam splitter with a squeezed
vacuum injected into the unused port. They have shown that the interference
pattern may be preserved if the input light is squeezed in a suitable
quadrature. As a model for phase-sensitive measurements,
Kennedy and Walls \cite{kn:kenw} suggested the use of a squeezed bath in place
of the thermal one, showing a substantial improvement of the
quadrature-phase sensitivity. On these lines, Bu\v{z}ek,
Kim and Gantsog \cite{kn:buz} \hfill have recently studied the phase
\newline \noindent \vskip 3.0truecm \noindent
properties of superpositions of coherent states evolving in
a squeezed (phase-sensitive) amplifier, showing that differently from
the phase-insensitive case \cite{kn:dami}, the phase distribution of the
Schr\"odinger-cat input state can be preserved for long times.

The above models either suffer a parametrically imposed dynamics for the
loss (beam splitter models\cite{kn:yusto,kn:mto,kn:tome,kn:meto}), or do
not consider the nonlinear coupling generating the superposition state,
which is just treated as an initial condition \cite{kn:kenw,kn:buz}. For this
reason we consider a model in which both the anharmonic Hamiltonian and the
interaction with a squeezed bath are taken  into account, thus allowing a test
of the effect of squeezed fluctuations on the generation of a superposition
state starting from a single coherent one.

This article is organized as follows: in Sec.~\ref{model}
we introduce the model and discuss the corresponding master equation.
In Sec.~\ref{results} we present the numerical integration of the
master equation
and the results of the numerical analysis, showing the efficiency
of a squeezed bath in the creation of a ``quasi-superposition state''.
In Sec.~\ref{conclu} we conclude and summarize the results.

\section{The model and the master equation}
\label{model}

We consider a mode of the electromagnetic field at frequency $\omega$ in a
Kerr medium coupled to a squeezed bath, namely a reservoir of
oscillators whose fluctuations are squeezed \cite{kn:kenw}.
The total Hamiltonian is given by
\begin{equation}
H=H_{\rm S}+H_{\rm I} +H_{\rm B}\;,
\label{eq:tothami}
\end{equation}
where $H_{\rm S}$ is the free Kerr Hamiltonian of the field mode
\begin{equation}
H_{\rm S} = \omega(a^{\dagger}a)+\Omega(a^{\dagger}a)^2\;,
\label{eq:hamis}
\end{equation}
$H_{\rm B}$ is the free Hamiltonian of the bath, and $H_{\rm I}$ is
the oscillator-reservoir interaction Hamiltonian, which in the rotating wave
approximation has the form
\begin{equation}
H_{\rm I} = a^{\dagger}\hat{\Gamma}+a\hat{\Gamma}^{\dagger}\;.
\label{eq:hamint}
\end{equation}
In Eqs.~(\ref{eq:hamis}) and (\ref{eq:hamint})
$a^{\dagger}$ and $a$ are the boson creation and annihilation operators of the
mode, while $\hat{\Gamma}^{\dagger}$ and $\hat{\Gamma}$ are bath operators.
The bath is squeezed and Markovian, namely the correlations functions of the
operators $\hat{\Gamma}^{\dagger}$ and $\hat{\Gamma}$ are given
by~\cite{kn:gard}
\begin{eqnarray}
\langle\hat{\Gamma}^{\dagger}(t)\hat{\Gamma}(t')\rangle & = & 2\gamma
N\delta(t-t')\;,
\nonumber \\
\langle\hat{\Gamma}(t)\hat{\Gamma}^{\dagger}(t')\rangle & = &
2\gamma(N+1)\delta(t-t')\;,
\nonumber \\
\langle\hat{\Gamma}(t)\hat{\Gamma}(t')\rangle & = & 2\gamma
Me^{-2i\omega t}\delta(t-t')\;,
\nonumber \\
\langle\hat{\Gamma}^{\dagger}(t)\hat{\Gamma}^{\dagger}(t')\rangle & = & 2\gamma
M^{\ast}e^{2i\omega t}
\delta(t-t')\;.
\label{eq:corr}
\end{eqnarray}
In Eqs. (\ref{eq:corr}) $\gamma$ is the damping constant (determined by the
coupling between the oscillator and the bath), $N$ is a real parameter,
which reduces to the mean number of thermal photons when the bath is not
squeezed, and $M$ is the squeezing complex parameter ($M=|M|e^{i\psi}$)
satisfying the relation
\begin{equation}
|M|\le \sqrt{N(N+1)}\;.
\label{eq:emme}
\end{equation}
The parameters $N$ and $M$ measure the strength of the correlations
of the bath degrees of freedom: for $|M|^2=N(N+1)$ the squeezing is
maximum, whereas for $|M|=0$ the reservoir reduces to the customary thermal
one.

{}From Eqs.~(\ref{eq:tothami})--(\ref{eq:corr}) one can derive the
following master equation for the reduced density matrix of the field
mode in the interaction picture \cite{kn:gard}
\FL
\begin{eqnarray}
{d\hat{\rho}\over dt} & = & -i\Omega[(a^{\dagger} a)^2,\hat{\rho}]
+\gamma(N+1)(2a\hat{\rho} a^{\dagger}-a^{\dagger} a\hat{\rho}-\hat{\rho}
a^{\dagger} a) \nonumber \\
 & & \mbox{}
+ \gamma N(2a^{\dagger}\hat{\rho} a-aa^{\dagger}\hat{\rho}-\hat{\rho}
aa^{\dagger}) \nonumber \\
 & & \mbox{}
- \gamma M(2a^{\dagger}\hat{\rho} a^{\dagger}-a^{\dagger}
a^{\dagger}\hat{\rho}-\hat{\rho} a^{\dagger} a^{\dagger}) \nonumber \\
 & & \mbox{} -\gamma M^{\ast}(2a\hat{\rho} a-aa\hat{\rho}-\hat{\rho}
aa)\;.
\label{eq:meq}
\end{eqnarray}
Eq. (\ref{eq:meq}) is numerically solved in the next section. The results are
shown in terms of the $Q$-function
\begin{equation}
Q(\alpha,\alpha^{\ast},t)=\langle\alpha |\hat\rho(t) |\alpha\rangle\;,
\label{eq:qu}
\end{equation}
which is the (positive definite) probability density for the
antinormally ordered moments of the annihilation and creation operators,
and in terms of the Wigner function, which is defined by
\begin{equation}
W(\alpha,\alpha^{\ast},t)=\int{{d^2\lambda}\over{\pi^2}}e^{-\lambda\alpha
^{\ast}+{\lambda}^{\ast}\alpha }\,\hbox{Tr}\left\{\hat\rho (t)e^{\lambda
a^{\dagger}-{\lambda^{\ast}}a} \right\} \;.
\label{eq:wigne}
\end{equation}

Using standard methods \cite{kn:miw}, it is possible to convert
the master equation~(\ref{eq:meq}) into the Fokker-Planck type equation for
the $Q$-function
\FL
\begin{eqnarray}
{{\partial Q}\over{\partial t}} & = & -i\Omega\left[
\alpha^{\ast}(1+2|\alpha|^2){{\partial Q} \over {\partial \alpha^{\ast}}}
-\alpha(1+2|\alpha|^2){{\partial Q}\over{\partial\alpha}}\right. \nonumber \\
 & & \mbox{} \; \; \; \; \; \; \; \;
+\left.(\alpha^{\ast})^2{{\partial^2Q}\over{\partial{\alpha^{\ast}}^2}}
-\alpha^2{{\partial^2Q}\over{\partial\alpha^2}}\right] \nonumber \\
 & & \mbox{} +\gamma\left({{\partial}\over{\partial\alpha}}
\alpha+{{\partial}\over{\partial\alpha^{\ast}}}\alpha^{\ast}\right)Q
+2\gamma(N+1){{\partial^2Q}\over{\partial\alpha\partial\alpha^{\ast}}}
\nonumber \\
 & & \mbox{} +\gamma M{{\partial^2Q}\over{\partial\alpha^2}}
+\gamma M^{\ast}{{\partial^2Q}\over{\partial{\alpha^{\ast}}^2}}\;.
\label{eq:fp}
\end{eqnarray}
Eq. (\ref{eq:fp}) has been solved for $\Omega=0$ \cite{kn:kenw} and for $M=0$
\cite{kn:dami}. For $\Omega$ and $M$ both nonvanishing, however, an
analytic solution is not available, whereas numerical integration has to
be carried out carefully, because of computation instabilities \cite{kn:colla}.

In absence of dissipation an initial coherent state
\begin{equation}
|\alpha_0\rangle = \exp\left(-{|\alpha_0|^2\over 2}\right)
\sum_{n=0}^{\infty}{\alpha_{0}^{n}\over\sqrt{n!}}|n\rangle
\label{eq:coe}
\end{equation}
evolves towards the following superposition of coherent states at $t=\pi
/2\Omega$
\begin{equation}
|\phi\rangle = {1\over\sqrt{2}}\left(e^{-i\pi/4}|\alpha_0\rangle +
e^{i\pi/4}|-\alpha_0\rangle\right)\;.
\label{eq:yusto}
\end{equation}
The $Q$-function of the $t=0$ coherent state is the Gaussian
\begin{equation}
Q(\alpha,\alpha^{\ast},0)=\exp(-|\alpha-\alpha_0|^2)\;,
\label{eq:incon}
\end{equation}
whereas for $t=\pi/2\Omega$ it approximatively corresponds to two Gaussian
peaks in the complex plane centered at $\alpha_0$ and $-\alpha_0$. The two
peaks are \lq\lq macroscopically distinguishable\rq\rq\ for
$|\alpha_0|\gg 1$. The evolution is periodic with period
$T=2\pi/\Omega$.
In presence of dissipation the state at times $t>0$ is
no longer pure, and a superposition of states corresponds to a \lq\lq
quasi-superposition\rq\rq\ mixed state, with interfering features still
surviving. We will look for signatures of such states from peaks in the
complex plane approximately separated by a distance $2|\alpha_0|$,
reminiscent of the state (\ref{eq:yusto}).

\section{Numerical results}
\label{results}

We have studied the time evolution of the density matrix by integrating
numerically the master equation (\ref{eq:meq}) for truncated Hilbert space
dimension $d=128$ (typically a power of 2, in order to take advantage of fast
Fourier transform algorithm for reconstructing the $Q$-function and the
Wigner function). The
integration time-step has to be carefully tuned as a function of $\Omega$
and $\alpha_0$ (here typically $\gamma\Delta t\simeq 10^{-5}$ for a
standard fourth-order Runge-Kutta routine). Numerical accuracy is
checked through normalization of $\hat\rho$, $Q$ and $W$, positivity of $Q$,
and reality of the diagonal elements of $\hat\rho$. As a test, the
results from analytical solutions for $\gamma =0$
\cite{kn:yusto,kn:milb} and for $\Omega=0$ \cite{kn:kenw} have been
recovered up to the seventh digit.

In Fig.~\ref{fg:sque} the
$Q$-function from the master equation (\ref{eq:meq}) for maximally
squeezed bath [$|M|=\sqrt{N(N+1)}$] is given for $\Omega   /\gamma =10$,
$N=30$, and squeezing phase $\psi=0$. At $\gamma t=0.0576$ [well before
$\Omega  t=\pi /2$, namely before the creation time of the state
(\ref{eq:yusto}) in the undamped case] two peaks are clearly visible in
the structure of the $Q$-function. This result can be compared with that
of Daniel and Milburn \cite{kn:dami} at the same time, but in absence of
squeezing. The two peaks are almost symmetrical with respect to the
origin of the phase space. Of course, due to
dissipation, they are not strictly coherent, but they are still visible,
even for subsequent times.

In Fig.~\ref{fg:term} the effect of a purely thermal bath on the
generation of the superposition state is shown for comparison with Fig.
\ref{fg:sque} for the same values of the parameters. In this case there
are no peaks in the $Q$-function, which now exhibits a classical
behaviour \cite{kn:miho}: this is the usual effect of
dissipation \cite{kn:caleg}.

In Fig.~\ref{fg:wign} we show for completeness the time evolution
of the Wigner function at the same times as in Fig.~\ref{fg:sque}:
the two peaks still show up, but now some interference features
arise between the two component states, as expected.
We have also considered the case in which the phase of the complex squeezing
parameter $M$ is chosen as $\psi=\pi$, namely the squeezing direction
of the phase sensitive bath is rotated of $90^{\circ}$ with respect to the
case $\psi=0$\cite{kn:buz,kn:loud}: the creation of two peaks
is still visible and their distance is approximately the same as
in Fig.~\ref{fg:sque}, but the line connecting them is now rotated of about
$90^{\circ}$, in agreement with the phase of $M$.

The effect is further confirmed by the behaviour of the quadrature
distributions $P(x_{\phi})$, defined by
\begin{equation}
P(x_{\phi})=\langle x_{\phi}|\hat{\rho}| x_{\phi}\rangle\;,
\label{eq:margi}
\end{equation}
with
\begin{equation}
\hat{x}_{\phi}=\frac{1}{\sqrt{2}}(e^{i\phi}a+e^{-i\phi}a^{\dagger})\;.
\label{eq:xfi}
\end{equation}
The probability $P(x_{\phi})$ is plotted in Fig.~\ref{fg:marg}
at $\gamma t=0.0576$ for $\phi=\pi/4$ and $\phi=3\pi/4$. For
$\phi=3\pi/4$---along the direction which joins the two peaks of the Wigner
function---the probability $P(x_{\phi})$ exhibits two peaks resulting
from marginal integration in the complex plane. On the other hand, along
the orthogonal direction at $\phi=\pi/4$, a pattern reminiscent of
interference in phase space\cite{kn:inte} between the two originally coherent
components is visible. We have also computed the degree of mixing
$S=1-\mathop{Tr}(\hat{\rho}^2)$: its behaviour shows a rapid increase
with time, achieving the asymptotic value $S=0.96$. Thus, the above coherence
features still survive notwithstanding the large mixing. Notice that,
however, this happens only for the two-peak state, whereas there is no
reminiscence of the symmetrical superposition of three coherent states
which arises at $t=\pi/3\Omega$ in absence of dissipation.

As regards the other parameters, the overall scenario is not affected by
changing the phase of the initial coherent field $\alpha _0$, whereas
increasing the modulus of $\alpha _0$ deteriorates the
formation of the two peaks. On the other hand,
slightly increasing $\Omega$ or $N$ (still for a maximally squeezed
bath), does not appreciably improve the visibility of the peaks.

\section{Summary and conclusions}
\label{conclu}

We have considered the anharmonic oscillator interacting with a
squeezed bath. The dissipative Kerr effect models a coherent state
propagating into a nonlinear medium, as, for example, into an optical
fiber. The squeezing has been considered in order to have the production
of coherent effects in presence of dissipation. Differently from the
previous works \cite{kn:kenw} and \cite{kn:buz}, our model treats all
effects (nonlinearity, dissipation and squeezing) contextually. We have
shown that squeezing the fluctuations of the bath improves the
generation of \lq\lq quasi-superposition\rq\rq\ states, which are
completely forbidden in presence of a heat bath. Our results are in
agreement with those of Kennedy and Walls \cite{kn:kenw} and of
Bu\v{z}ek, Kim and Gantsog \cite{kn:buz} and complete them: a squeezed
bath not only is able to preserve a macroscopic quantum superposition, but
also allows its generation in presence of nonlinear interactions.
The physical meaning of a squeezed bath is not completely understood in
the present context: a suitable feedback mechanism could be envisaged
that supports an {\em ad hoc} phase sensitive interaction, which
amplifies or attenuates fluctuations, depending on the
quadrature \cite{kn:tovi}.

In conclusion, some comments regarding the direction of squeezing in the bath
are in order. Despite the squeezing direction does not affects the generation
of quasi-superposition states, the survival time of coherence
is naturally increased only for squeezing in the direction orthogonal to the
line that joins the component states. On the other
hand, the Kerr effect rotates the $Q$-function in the complex plane,
thus making squeezing less efficient in the overall evolution. This
suggests that an ideal bath should be squeezed isotropically in the
plane, depending on the phase of the state itself. It is possible to
write a master equation for such an isotropically squeezed bath:
numerical results on these lines will be the object of a forthcoming
paper \cite{kn:dafot}.

We would like to thank Prof. M. Cini for many stimulating and
helpful discussions on this topic.

\begin{figure}
\caption{Contour plots of the time evolved $Q$-function according
      to Eq.~(\protect\ref{eq:meq}). Here $N=30$, $|M|=\protect\sqrt{N(N+1)}$,
        $\psi=0$, $\alpha_0 =3$, and $\Omega/\gamma=10$.}
\label{fg:sque}
\end{figure}

\begin{figure}
\caption{The same as in Fig.~\protect\ref{fg:sque}, but for $M=0$.}
\label{fg:term}
\end{figure}

\begin{figure}
\caption{The Wigner function corresponding to plots in
Fig.~\protect\ref{fg:sque}.}
\label{fg:wign}
\end{figure}

\begin{figure}
\caption{Plot of the marginal distribution $P(x_{\phi})$ at $\gamma
t=0.0576$ for $\phi=3\pi/4$ and $\phi=\pi/4$.}
\label{fg:marg}
\end{figure}

\end{document}